\documentclass{llncs_Ibergrid2009}
\usepackage[dvipdf]{graphicx}
\usepackage{hyperref}
\usepackage{color}
\usepackage[utf8x]{inputenc}
\newcommand{\arXiv}[1]{\href{http://arxiv.org/abs/#1}{\tt arXiv:#1}}

\def\aitalc{{\sc aitalc}}
\def\form{{\sc form}}
\def\fortran{{\sc fortran}}
\def\diana{{\sc diana}}
\def\gridway{{\sc GridWay}}
\def\looptools{{\sc looptools}}
\def\cpp{{\sc c}{\footnotesize ++}}

\begin{document}

\frontmatter
\pagestyle{headings}  % switches on printing of running heads

\mainmatter              % start of the contributions
%\title{Porting AITALC product through a master-worker scheme}
\title{Grid porting of Bhabha scattering code through a master-worker scheme}
\author{Alejandro Lorca, José Luis Vázquez-Poletti, Eduardo Huedo,\\ Ignacio M. Llorente}
\institute{Distributed System Architecture group,\\Dpto. de Arquitectura de Computadores y Automática\\Facultad de Informática, Universidad Complutense de Madrid,\\C/ Prof.\ José García Santesmases s/n, E-28040 Madrid, Spain.\\
\email{alejandro.lorca@fdi.ucm.es}, \email{jlvazquez@fdi.ucm.es}, \email{ehuedo@fdi.ucm.es},\\ \email{llorente@dacya.ucm.es}
}

\maketitle

\begin{abstract}
A program calculating Bhabha scattering at high energy colliders is considered for porting to the EGEE Grid infrastructure. The program code, which is a result of the \aitalc{} project, is ported by using a master-worker operating scheme. The job submission, execution and monitoring are implemented using the \gridway{} metascheduler. The unattended execution of jobs turned out to be complete and rather efficient, even when pre-knowledge of the grid is absent. While the batch of jobs remains organized at the user's side, the actual computation was carried out within the phenogrid virtual organization. The scientific results support the use of the small angle Bhabha scattering for the luminosity measurements of the International Linear Collider project.
\end{abstract}

\section{Introduction}
\label{introduction}
The International Linear Collider (ILC) is an electron-positron accelerator planned to supersede the Large Hadron Collider (LHC) and lead the high energy physics research in the nearby decades. Still, the proposal and supporting groups for the ILC are awaiting the first signals of LHC to complement their goals to the expected discoveries from the proton-proton collider. Thousands of scientists and engineers firmly believe \cite{:2007sg,Djouadi:2007ik} on the advantage coming from a cleaner environment 
 resulting from the electron-positron collisions at the ILC. This translates into a clearer signal to background ratio than in the LHC due to the absence of plethora of hadronic subproducts.
Therefore, this kind of accelerator could much precisely determine important parameters of the Standard Model of particle physics and thus lead to tighter model constraints and even discoveries ``by precision''. Some high-precision measurements in history, like planetary motion, the speed of light or, more recently, deep inelastic scattering, led to remarkable advances in our understanding of astronomy, special relativity and quantum chromodynamics (QCD) respectively. They remind us the importance of such methodology.

The calculations involved in an accurate determination of any physical observable are, within perturbation theory in quantum field theory, rather cumbersome. Fortunately, computing resources have grown fast enough during the last decades to accomplish these calculations despite of their increasing complexity. Automated software tools have arosen in the last thirty years to provide systematic and reliable answer for these predictions.

In this context, the EGEE project\footnote{Enabling Grids for E-sciencE, INFSO-RI-222667, 7th Framework Programme. {\tt http://www.eu-egee.org/}.} provides a computing infrastructure where scientists and engineers perform numerous studies and tests in order to provide an efficient distributed architecture for Grid computing. 

Along this paper we describe the scientific goals (Sec.~\ref{scientific}) and the methodology we used for adapting a code for theoretical predictions to run properly on to the Grid (Sec.~\ref{porting}). Analyses on the execution of jobs (Sec.~\ref{execution}) and scientific results (Sec.~\ref{results}) base the case study. Finally a look at further possibilities and conclusions finishes this contribution on Sec.~\ref{conclusions}.

\section{Scientific scope}
\label{scientific}
Nowadays, high energy physics applications intensively use Grid resources during data processing and analyses from LHC. Theoreticians also require heavy computational tasks to match the same level of accuracy achieved by the experimental measurements. In a close future, after the LHC experience, it could be very well stablished that Grid technologies are a standard procedure in order to achieve the permill level of uncertainty expected for the ILC.

In this article we will consider Bhabha scattering (i.e., the reaction of an electron-positron pair into themselves)
\begin{equation}
e^- e^+ \rightarrow e^- e^+,
\end{equation}
 as our target process.

One of the reasons to consider Bhabha scattering is the high cross section resulting at small angle due to the small deflection coming mainly from electromagnetic interactions. The resulting large amount of statistics within a well understood model allows for luminosity calibration. Such technique has been used at high energy colliders like LEP and SLD. Additionally, large angle Bhabha scattering has been also used at low energy $b$-quark factories like Belle or BaBar.

On the computational side, \aitalc{} \cite{Lorca:2004fg} is a useful tool for automating some calculations needed by theoreticians and phenomenologists in high energy physics. It produces numerical programs directly from the symbolic representation of the process, by using the so-called Feynman rules \cite{Feynman:1948ur} and generating the complete analytic expressions for the respective Feynman diagrams within perturbation theory. Since the package is able to produce code for scattering of 2 $\to$ 2 fermions, it was suited to deliver the necessary routines to compute integrated cross sections for Bhabha scattering, including the first order corrections coming from electroweak quantum loop effects and soft-photon radiation.

The goal was to scan the full range of centre-of-mass energies, from the 10 GeV at the B-factories, to the 1000 GeV expected to be achieved at the second running phase of the ILC. This range will be covered by 2048 data-points from both, tree-level (zeroth order) and first order corrected integrated cross sections. This coverage permits to quantify the impact of the quantum loops effects and relays in an complementary simulation of hard-photon effects \cite{Berends:1973jb} by Monte Carlo programs \cite{Jadach:1991by}. Moreover two-loop photonic corrections \cite{Bonciani:2004qt,Czakon:2004wm,Penin:2005kf} also play a role in order to complete the full theoretical prediction.

\section{Porting the application}
\label{porting}
\subsection{Preliminaries}
Whilst \aitalc{} is a tool integrating three other independent packages; \form{} \cite{Vermaseren:2000nd}, \diana{} \cite{Tentyukov:1999is} and \looptools{} \cite{Hahn:1998yk}, the resulting code which brings the numerics to the end-user is built as a dynamically linked \fortran{} executable. 

Our primary intention was to port to the Grid the complete process, the generating tool and the code execution.

\begin{figure}[!ht]
\begin{center}
\includegraphics[scale=.4]{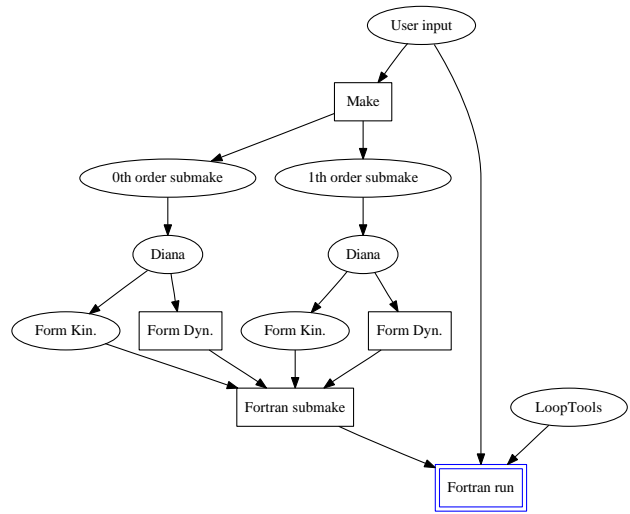}
\caption{Job work-flow for a typical process study with \aitalc{}. Squares enclose those tasks with an internal structure suitable for some kind of parallelization. Double squared box shows the most suitable part to be ported to the Grid.}
\label{workflow}
\end{center}
\end{figure}

Fig.~\ref{workflow} describes the work-flow of the tool \aitalc{}. There is a main make process which organizes the tree-level (0th order submake), the loop corrections (1st order submake) and the last numerical (\fortran{} submake) part through automated Makefiles. These three processes depend on each other as depicted, so the 0th and 1st order could be run in parallel. Because the complexity of the internal calculations, it turns out that the 1st order demands much more resources than the 0th order. Still there are some internal parts of the branches, the dynamical ones under \form{} (\form{} Dyn.), which could in principle also be parallelized. The last block which evaluates the numerics (\fortran{} run) strongly depends on the user's needs.

There are two main reasons why we considered only the final task to be ported to the Grid:
\begin{itemize}
\item {\bf Independence}. \aitalc{} runs natively on Unix machines, but even if it is not uncommon to find dedicated machines hosting \fortran{} compilers, \form{}, \diana{} and \looptools{}, are very specialized software packages whose licences and distribution channels avoid out-of-the-box availability in standard Linux distributions. For this reason, \aitalc{} supplies an installation script sorting out these inconveniences, but is unfeasible to provide a single executable and the whole tool with these three packages should be installed altogether in each running node. The later scenario immediately conflicts with standardized user permissions and node specifications.
\item {\bf Timings}. The production of the \fortran{} code does not take much time in a single machine for an example process. Our full-massive electroweak Bhabha scattering, being the largest process \aitalc{} is able to generate, took not much more in our testing machine, just a few minutes to finish as shown in Tab.~\ref{timings_aitalc}. Because the total time to evaluate 2048 data-points is approximately two orders of magnitude larger (e.g. Tabs.~\ref{4_512}-\ref{32_64}), we might neglect the time to create the \fortran{} code.
\end{itemize}
\begin{table}[!ht]
\begin{center}
\begin{tabular}{|l|r|}
\hline
{Task}&{Time [s]}\\
\hline
0th order submake & 12\\
1st order submake & 381\\
\fortran{} submake & 57\\
\hline
\end{tabular}
\end{center}
\centering
\caption{Detailed running time for each building block of \aitalc{} when producing the \fortran{} code for Bhabha scattering.}
\label{timings_aitalc}
\end{table}

\subsection{Master-worker scheme}
\begin{figure}[!ht]
\begin{center}
\includegraphics[scale=.45]{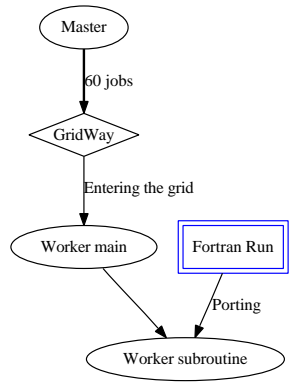}
\caption{Grid job work-flow. The \gridway{} metascheduler acts as middleware stablishing communication between master and worker.}
\label{workflow_ported}
\end{center}
\end{figure}
The running profile of the application let the {\em master-worker} scheme being a very suitable one in order to achieve a balance between user intervention and execution performance. A scheme of the final application is depicted in the  Fig.~\ref{workflow_ported}. It shows the work-flow consisting of a major master piece which decides how heavy the jobs are going to be (according to the user's instructions) and a worker part being transferred and remotely run. The master communicates with the Grid via the {\tt gwsubmit} command of the \gridway{} metascheduler \cite{Huedo:2005}. \gridway{} metascheduler offers a clean and user-friendly interface to submit jobs into a Grid middleware like {\sc globus} or {\sc gLite}.

Being \gridway{} capable to perform unattended job migration, recovery and rescheduling \cite{Vazquez:2007}, it was extremely useful for getting, with the generic and simplest configuration, a complete set of jobs being delivered for execution at different computing elements. Thus a deeper study of the queue and hardware characteristics of each cluster is not required.

Converting the \fortran{} program into a worker was a minor issue which is nevertheless worth to be mentioned since it is a common problem met by the scientific community. We chose to subdivide the worker into two parts:
\begin{itemize}
\item A simple main program, written in \cpp{}, accepting running parameters as arguments passed through command line interface. This main program calls, for each parameter configuration, a unique \fortran{} subroutine.
\item The \fortran{} subroutine, which is an adaptation of the unported main program commenting out the settings for initial parameters. These are, according to \aitalc{} description: {\tt nsqrtsman}, {\tt minsqrtsman}, {\tt maxsqrtsman}, {\tt setlimcost}, {\tt setfracomega}.
\end{itemize}
Moreover, it was also required to ensure 32 bits compatibility and static linkage (usually flags {\tt -m32 --static} in compilers). In such a way, the final executable was suited to be run in any working node without lacking any external libraries. The size of this executable (5.8MB) did not create transfer time bottlenecks.

\section{Execution analysis}
\label{execution}
The scanning of data-points were planned in batches of jobs. Each of them contained specific instructions to generate automatically job templates with specific configuration for every job. This configuration had to do only with the definition of parameter range. The batches had equal amount of data to process, so we can consider the blocks equivalents to one another:
\begin{itemize}
\item 1st block: 4 jobs containing 512 data-points each,
\item 2nd block: 8 jobs containing 256 data-points each,
\item 3rd block: 16 jobs containing 128 data-points each,
\item 4th block: 32 jobs containing 64 data-points each.
\end{itemize}

The evolution of a successful job is as follows: First, after a computing element has been chosen and accepted, the waiting time for a working node is denoted as {\em pending}. There is some {\em stage-in} time required to transfer data to the computing element. Then, the job enters in its {\em execution} stage. Finally a {\em stage-out} time is needed to transfer back the output and status. Moreover we have also to add some {\em overhead} due to job request processing and state change notifications. Occasionally the job gets migrated after a {\tt suspension\_time} configured by default in \gridway{} to another computing element, in that case it appears in the graphics as {\em waiting}.
Regarding totals, we noted by {\em computing} time the whole time summed up for the different machines and with the keyword {\em human}, the waiting time from batch start to the end.

\begin{figure}[!th]
\begin{center}
\includegraphics[scale=0.5]{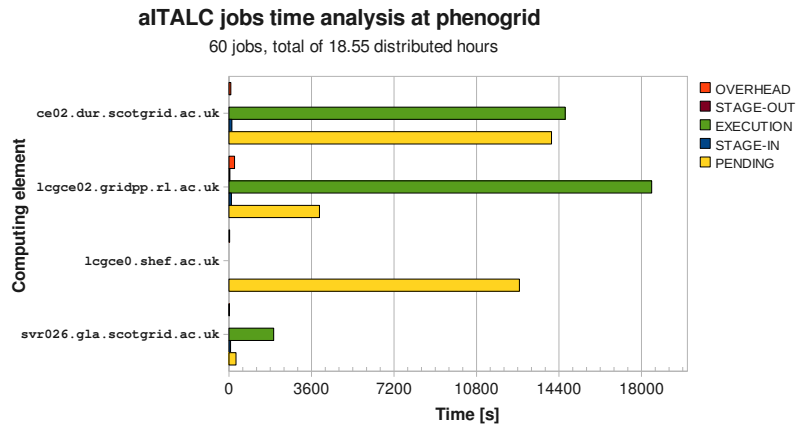}
\caption{Overall job-stage time for each computing element. The vertical order of bars is the same as shown in the legend.}
\label{timing_node}
\end{center}
\end{figure}

\begin{table}[!ht]
\begin{center}
\begin{tabular}{|l|r|r|r|r|r|r|r|r|}
\hline
 {\bf Computing Element} &\multicolumn{2}{|c|}{\bf CPU} & {\bf Nodes} & \multicolumn{5}{|c|}{\bf Time performance [s]}\\
\hline
FQDN & Type & MHz & Total & Pend. & St.-in & Exec. & St.-out& Overh.\\
%FQDN & Type & MHz & Total & Pending & Prolog & Active & Epilog& Prewrap \\
\hline
{ce02.dur.scotgrid.ac.uk}& Xeon & 2667& 672 &14074& 127&14670&0&78\\
{lcgce0.shef.ac.uk}&Opteron&2400&190 &12673&0&0&0&27\\
{lcgce02.gridpp.rl.ac.uk}&P.III& 1001&2525&3948& 104&18447&41&243\\
{svr026.gla.scotgrid.ac.uk} & Opteron& 2200& 1896&306& 66&1949&0&11\\
%ce00.hep.ph.ic.ac.uk&&&&&&&&\\
%ce01.dur.scotgrid.ac.uk&&&&&&&&\\
%ce01.esc.qmul.ac.uk&&&&&&&&\\
%ce02.tier2.hep.manchester.ac.uk&&&&&&&&\\
{ce1.pp.rhul.ac.uk}${}^*$&P.IV & 1000 & 136 &-&-&-&-&149\\
%ce2.ppgrid1.rhul.ac.uk&&&&&&&&\\
%ce.glite.ecdf.ed.ac.uk&&&&&&&&\\
%dgc-grid-40.brunel.ac.uk&&&&&&&&\\
%dgc-grid-44.brunel.ac.uk&&&&&&&&\\
%fal-pygrid-18.lancs.ac.uk&&&&&&&&\\
%fal-pygrid-44.lancs.ac.uk&&&&&&&&\\
%gw-4.ccc.ucl.ac.uk&&&&&&&&\\
%hepgrid2.ph.liv.ac.uk&&&&&&&&\\
%heplnx206.pp.rl.ac.uk&&&&&&&&\\
%heplnx207.pp.rl.ac.uk&&&&&&&&\\
%mw05.ecdf.ed.ac.uk&&&&&&&&\\
{svr021.gla.scotgrid.ac.uk}${}^*$&Opteron&1896&2200&-&-&-&-&18\\
%t2ce02.physics.ox.ac.uk&&&&&&&&\\
%FQDN & Type & MHz & Total & Stage-in & Overhead & Pending & Execution & Stage-out\\
%FQDN & Type & MHz & Total & Prolog & Prewrap & Pending & Active & Epilog\\
%\hline
%{ce02.dur.scotgrid.ac.uk}& Xeon & 2667& 672 & 127&78&14074&14670&0\\
%{lcgce0.shef.ac.uk}&Opteron&2400&190 &0&27&12673&0&0\\
%{lcgce02.gridpp.rl.ac.uk}&P.III& 1001&2525& 104&243&3948&18447&41\\
%{svr026.gla.scotgrid.ac.uk} & Opteron& 2200& 1896& 66&11&306&1949&0\\
%ce00.hep.ph.ic.ac.uk&&&&&&&\\
%ce01.dur.scotgrid.ac.uk&&&&&&&\\
%ce01.esc.qmul.ac.uk&&&&&&&\\
%ce02.tier2.hep.manchester.ac.uk&&&&&&&\\
%ce1.pp.rhul.ac.uk&P.IV & 1000 & 136 &-&149&-&-&-\\
%ce2.ppgrid1.rhul.ac.uk&&&&&&&\\
%ce.glite.ecdf.ed.ac.uk&&&&&&&\\
%dgc-grid-40.brunel.ac.uk&&&&&&&\\
%dgc-grid-44.brunel.ac.uk&&&&&&&\\
%fal-pygrid-18.lancs.ac.uk&&&&&&&\\
%fal-pygrid-44.lancs.ac.uk&&&&&&&\\
%gw-4.ccc.ucl.ac.uk&&&&&&&\\
%hepgrid2.ph.liv.ac.uk&&&&&&&\\
%heplnx206.pp.rl.ac.uk&&&&&&&\\
%heplnx207.pp.rl.ac.uk&&&&&&&\\
%mw05.ecdf.ed.ac.uk&&&&&&&\\
%svr021.gla.scotgrid.ac.uk&Opteron&1896&2200&-&18&-&-&-\\
%t2ce02.physics.ox.ac.uk&&&&&&&\\
\hline
\multicolumn{9}{|l|}{${}^*$ These computing elements reiteratively returned job-callback errors.}\\
\hline
\end{tabular}
\end{center}
\centering
\caption{Computing elements used for scheduling belonging to the {\tt pheno} Virtual Organization at the EGEE infrastructure. Sums of short times rounded up to a second may be shown as zero, even if there was some accumulated latency.}
\label{timings_node}
\end{table}
All the jobs were submitted to the infrastructure phenogrid \footnote{Phenogrid, particle physics phenomenology on the grid. {\tt http://www.phenogrid.dur.ac.uk/}.}, which takes part in the EGEE project under the phenomenology-focussed virtual organization {\tt pheno}.
The timing analysis by cluster is given in Tab.~\ref{timings_node} and graphically shown in Fig.~\ref{timing_node}. Three of the six computing elements worked properly and output data files were produced at regular rate. Other ({\tt lcgce0.shef.ac.uk}) seemed to have a misconfiguration \cite{david} and therefore left our jobs pending for too long and therefore being migrated automatically by \gridway{} to most capable hosts. With the other two computing elements ({\tt ce1.pp.rhul.ac.uk} and {\tt svr021.gla.scotgrid.ac.uk}) we found some errors after submitting our jobs that could be related to the local resource management system (LRMS), so \gridway{} applied a temporal banning policy to them to avoid unsuccessful retries. This is just one of the already implemented mechanisms to enhance performance without user intervention.

\begin{figure}[!t]
\begin{center}
\includegraphics[width=0.7\textwidth]{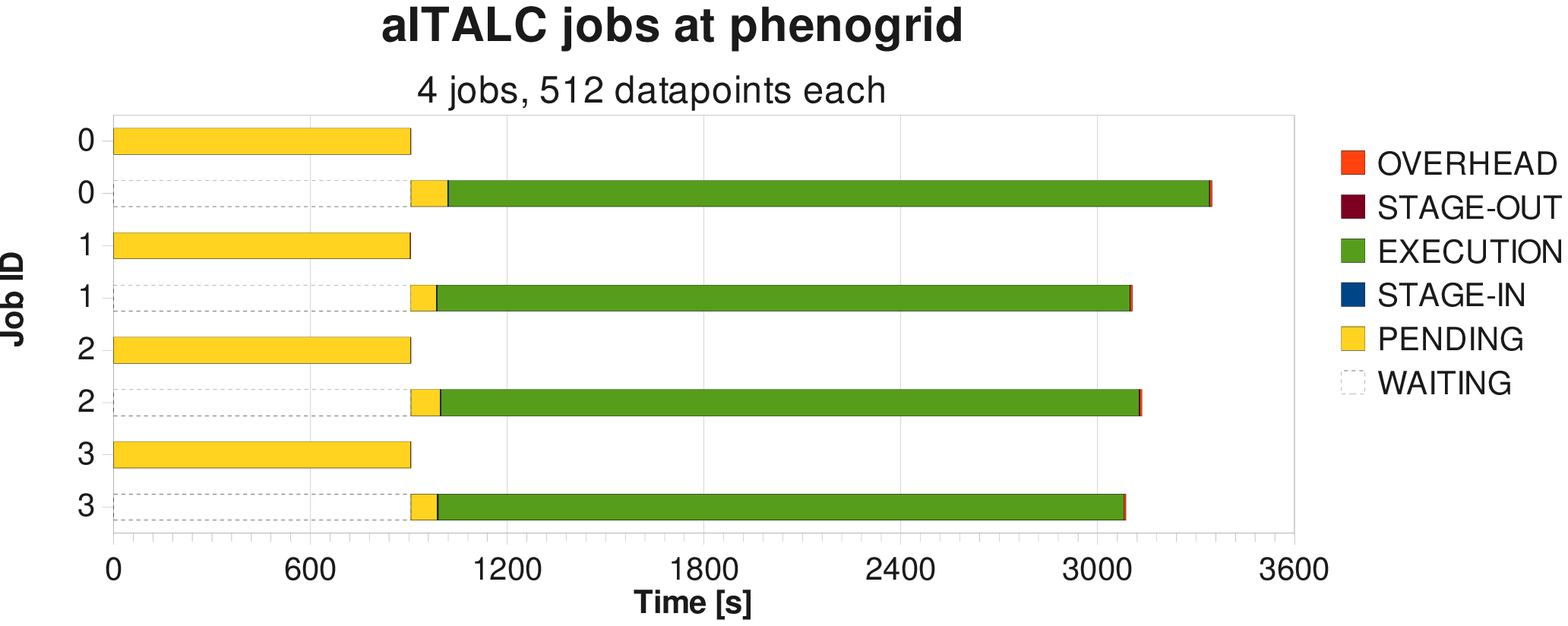}
\caption{Batch of 4 jobs computing 512 data-points each at phenogrid. The evolution of each job is split into separated bars every time it was migrated from computing element.}
\label{4_512}
\end{center}
\end{figure}

\begin{figure}[!t]
\begin{center}
\includegraphics[width=0.7\textwidth]{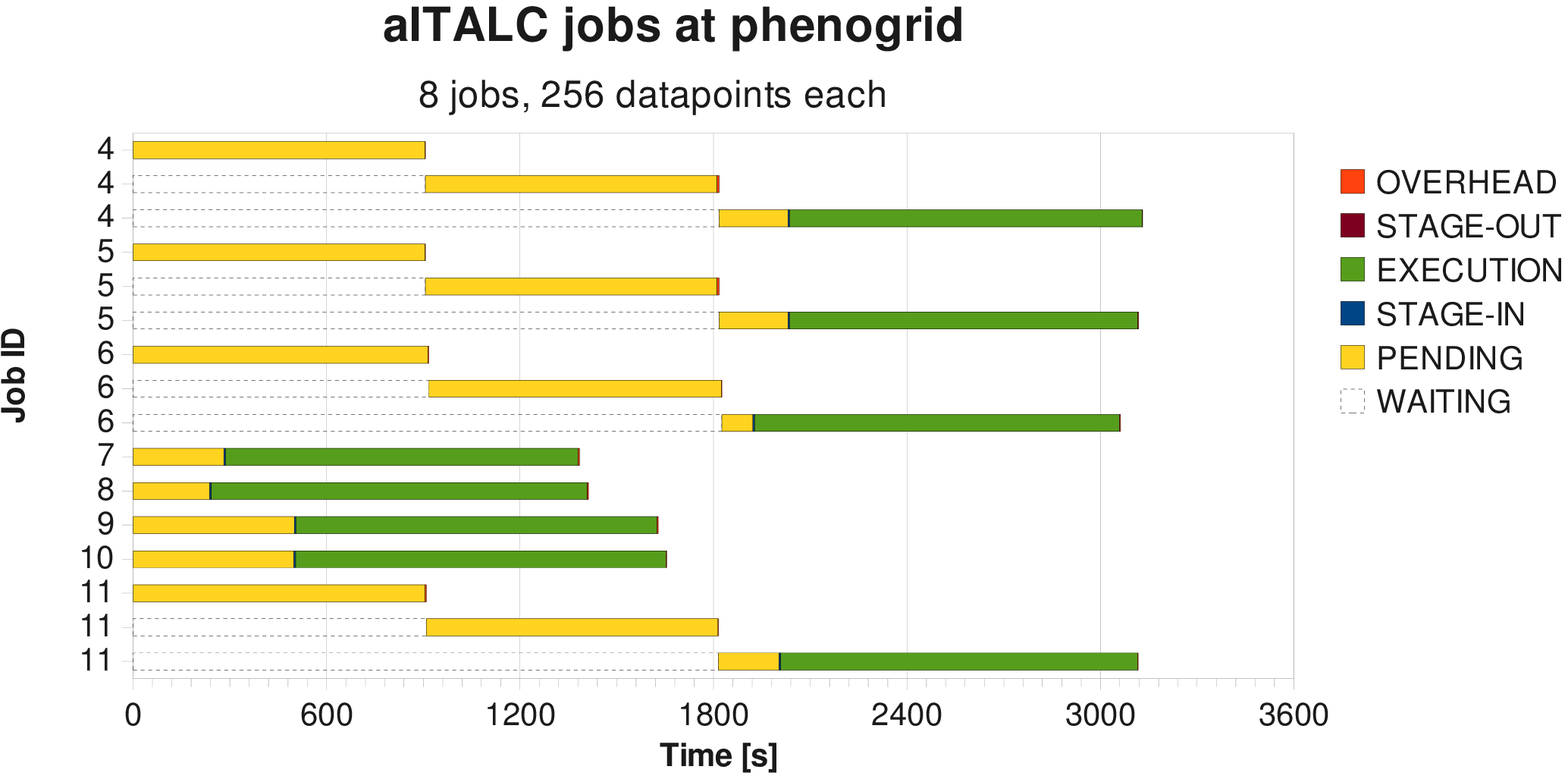}
\caption{Batch of 8 jobs computing 256 data-points each at phenogrid. The evolution of each job is split into separated bars every time it was migrated from computing element.}
\label{8_256}
\end{center}
\end{figure}

\begin{figure}[!t]
\begin{center}
\includegraphics[width=0.7\textwidth]{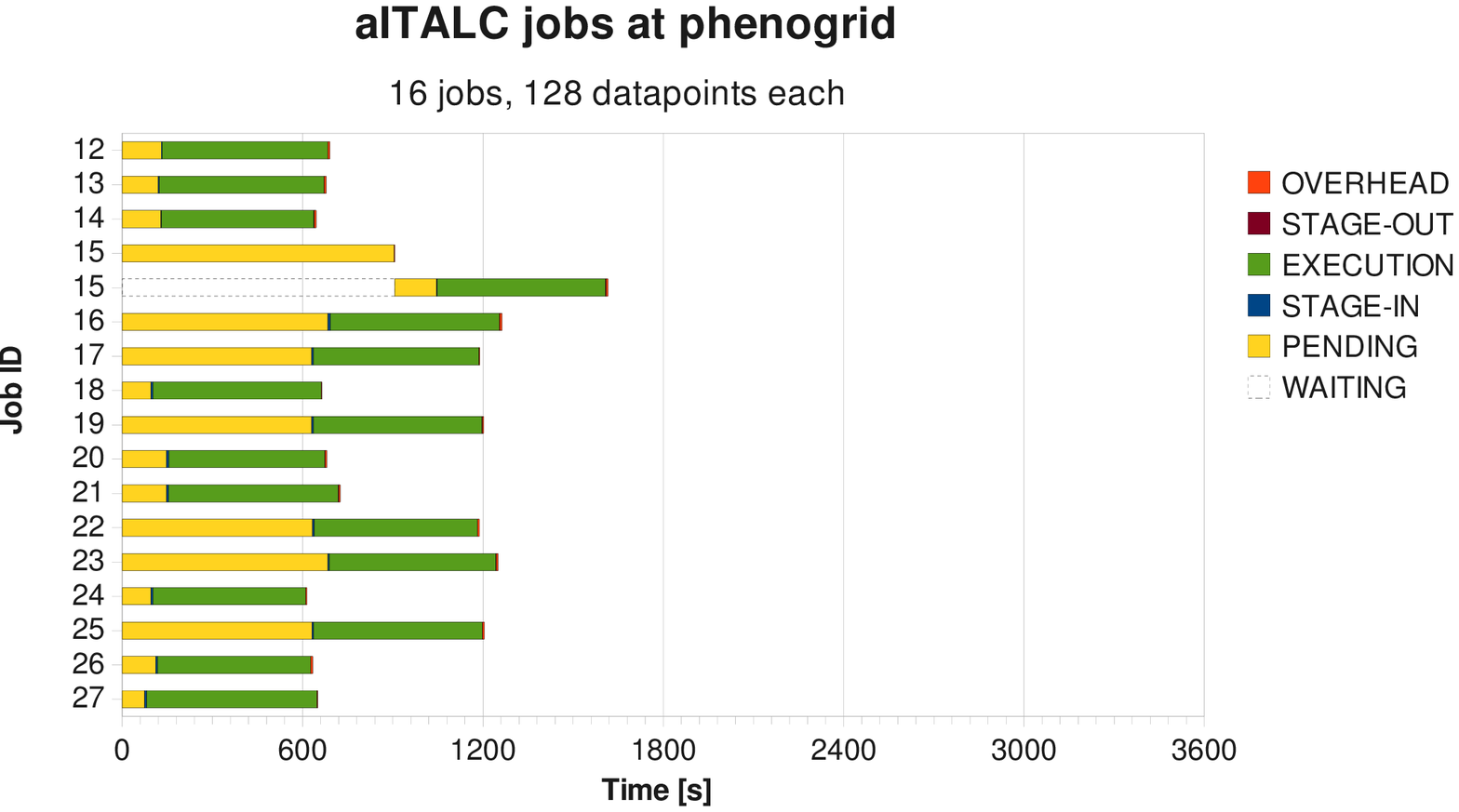}
\caption{Batch of 16 jobs computing 128 data-points each at phenogrid. The evolution of each job is split into separated bars every time it was migrated from computing element.}
\label{16_128}
\end{center}
\end{figure}

\begin{figure}[!ht]
\begin{center}
\includegraphics[width=0.7\textwidth]{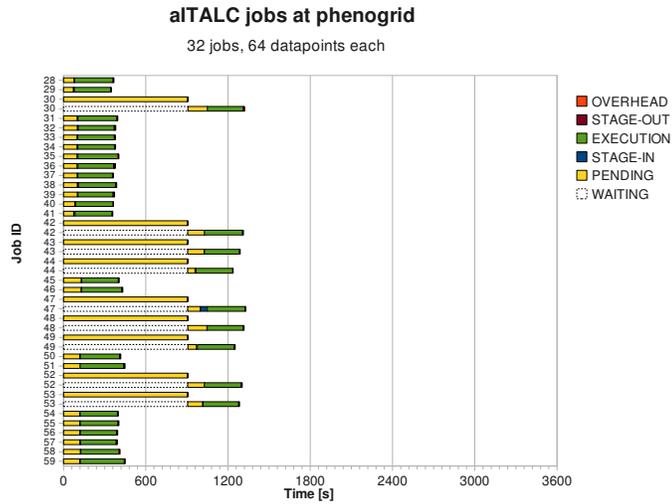}
\caption{Batch of 32 jobs computing 64 data-points each at phenogrid. The evolution of each job is split into separated bars every time it was migrated from computing element.}
\label{32_64}
\end{center}
\end{figure}

Having a look to Tab.~\ref{timings_block} the Figs.~\ref{4_512}-\ref{32_64}, we can appreciate the fair behaviour of the Grid computing. The more jobs contains the block, less {\em human} time waits the end-user, so parallelization works as expected. Nonetheless for a short amount of jobs as we have in this study, we cannot expect a linear statistical reduction due to the timeout of the pending time which some computing elements introduce. This timeout is configured in \gridway{} via the {\tt SUSPENSION\_TIMEOUT} parameter, and the end-user might include it in his job template. This setting, as well as many other performance enhance strategies, lie outside of this paper's scope.

\begin{table}[!t]
\begin{center}
\begin{tabular}{|l|r|r|r|r|r|r|}
\hline
{\bf Block} & \multicolumn{6}{|c|}{\bf Time performance [s]}\\
\hline
%Jobs $\times$ data-points & Prolog& Prewrap & Pending & Active & Epilog & Total\\
Jobs $\times$ data-points & Pending & Stage-in& Execution & Stage-out & Overhead & Total\\
\hline
\phantom{0}4 $\times$ 512 & 3980 & 11 & 8648 & 5 & 30 & 12674\\
\phantom{0}8 $\times$ 256 & 9484 & 42 & 8939 & 3 & 39 & 18507\\
16 $\times$ 128 & 5989 & 86& 8728 & 10 & 82 & 14895 \\
32 $\times$ \phantom{0}64 &11548& 158&8751& 23 & 208 & 20688\\
\hline
%& Pending/job & Prolog/job & Active/job & Epilog/job & Prewrap/job &  Human\\
& Pend./job & St.-in/job & Exec./job & St.-out/job & Overh./job &  Human\\
\hline
\phantom{0}4 $\times$ 512 & 995.00 & 2.75 & 2162.00 &1.25&7.50& 3348 \\
\phantom{0}8 $\times$ 256 & 1185.5 & 5.25 & 1117.38 & 0.38& 4.88 & 3131\\
16 $\times$ 128 &374.31 & 5.38&545.50 &0.63&5.13&1616\\
32 $\times$ \phantom{0}64 &360.88& 4.94 &273.47&0.72& 6.50 &1328\\
%Jobs $\times$ data-points & Stage-in& Overhead & Pending & Execution & Stage-out & Total\\
%\hline
%\phantom{0}4 $\times$ 512 & 11 & 30 & 3980 & 8648 & 5 & 12674\\
%\phantom{0}8 $\times$ 256 & 42 & 39 & 9484 & 8939 & 3 & 18507\\
%16 $\times$ 128 & 86& 82 & 5989 & 8728 & 10 & 14895 \\
%32 $\times$ \phantom{0}64 & 158& 208 &11548&8751& 23 & 20688\\
%\hline
%& Prolog/job & Prewrap/job & Pending/job & Active/job & Epilog/job &  Human\\
%& St.-in/job & Overh./job & Pend./job & Exec./job & St.-out/job &  Human\\
%\hline
%\phantom{0}4 $\times$ 512 & 2.75 &7.50& 995.00 & 2162.00 &1.25& 3348 \\
%\phantom{0}8 $\times$ 256 & 5.25 & 4.88 & 1185.5 & 1117.38 & 0.38& 3131\\
%16 $\times$ 128 & 5.38&5.13&374.31 &545.50 &0.63&1616\\
%32 $\times$ \phantom{0}64 & 4.94 & 6.50 &360.88&273.47&0.72&1328\\
\hline
\end{tabular}
\end{center}
\centering
\caption{Time performance by block distribution of jobs. Precision is limited by rounded up of invidual timings to the second.}
\label{timings_block}
\end{table}


\section{Scientific results}
\label{results}
The correct execution of all the jobs let us compose the complete scan of the integrated cross section for Bhabha scattering. Two scans were performed with crossed exponential stepping to ensure that a late/failed job could still be interpolated from the rest without much loss of precision. The following two configurations for different scattering angle $\theta$ and maximum soft photon energy $E^{\mathrm{max}}_{\gamma_{\mathrm{soft}}}$ were considered:
\begin{itemize}
\item Large angle: $-0.9 < \cos{\theta} < 0.9, E^{\mathrm{max}}_{\gamma_{\mathrm{soft}}} = 0.1 \sqrt{s}$
\item Small angle: $25~\textrm{mrad} < \theta < 90~\textrm{mrad, } E^{\mathrm{max}}_{\gamma_{\mathrm{soft}}} = 0.2 \sqrt{s}$
\end{itemize}
being $\sqrt{s}$ the centre-of-mass energy.

\begin{figure}[!t]
\begin{center}
\includegraphics[width=0.6\textwidth]{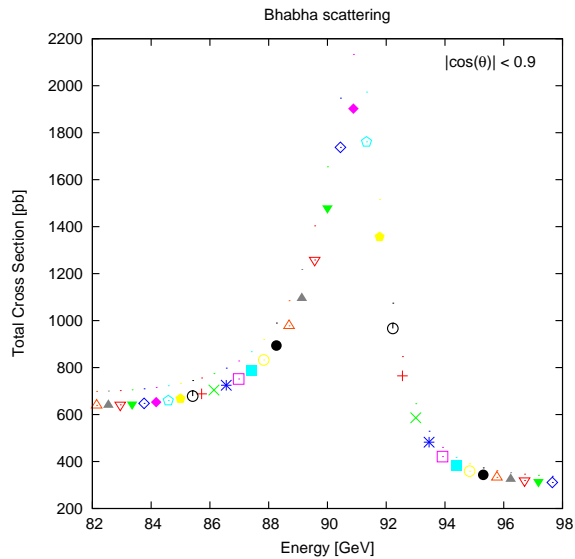}
\end{center}
\centering
\caption{Bhabha scattering around the $Z$-pole. Tiny dots stand for the tree-level while shaped symbols do it for the electroweak 1-loop plus soft photon corrections with $E^\mathrm{max}_{\gamma_{\mathrm{soft}}}=0.1 \sqrt{s}$. Each of the sixteen different symbols matches a different job identifier, being therefore the scanning of data-points computed in a distributed environment.}
\label{bhabha_09}
\end{figure}

We can observe the resonance induced by the gauge neutral $Z$-boson in Fig.~\ref{bhabha_09}. Here different symbols indicate different job identifiers, giving us therefore the complete result only when all the jobs are finished. The electroweak corrections are in this range quite important at percent level, since the angular cross section is stopped at the angles close to collinearity.

\begin{figure}[!ht]
\begin{tabular}{r}
\includegraphics[scale=0.64]{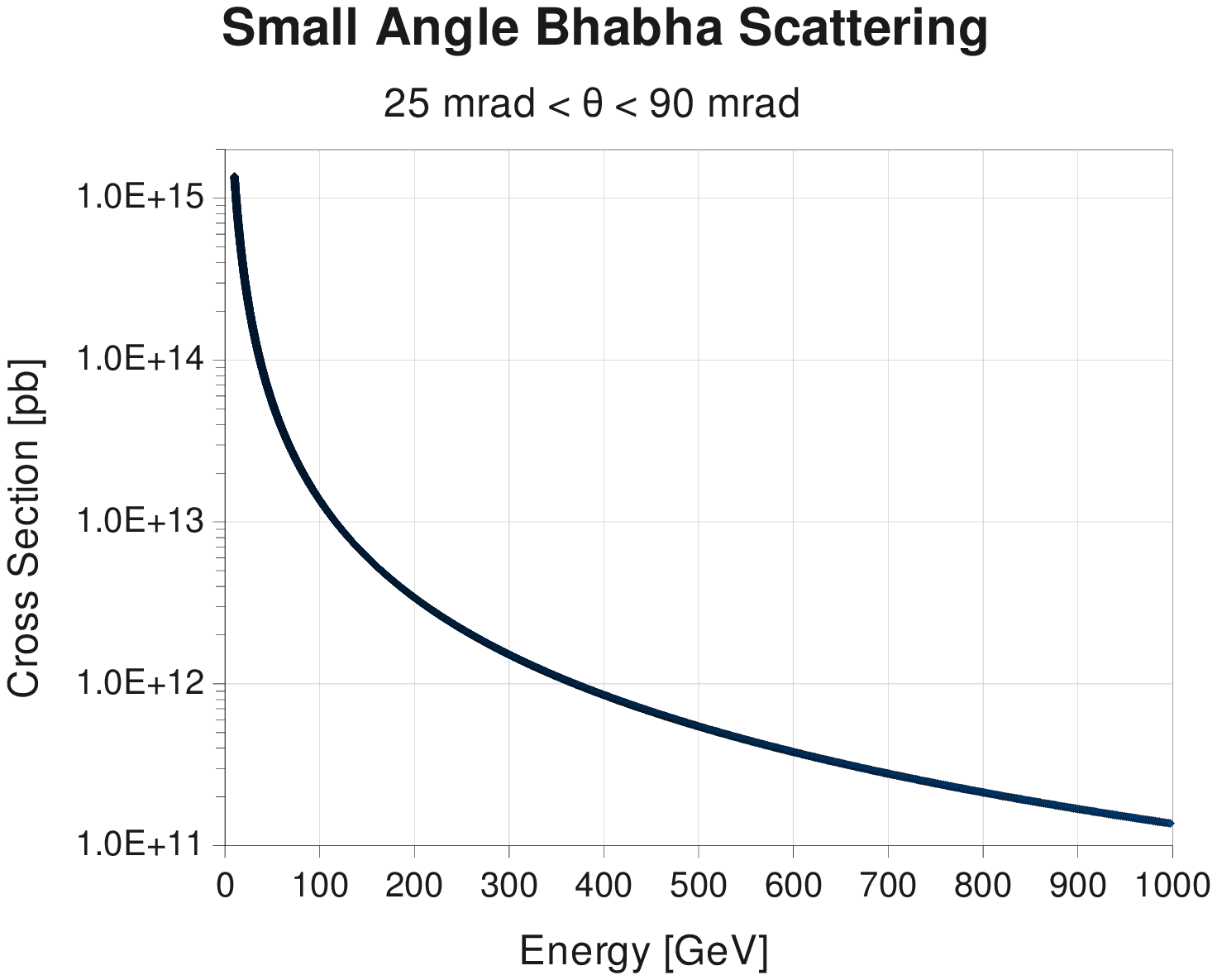}\\
\includegraphics[scale=0.6]{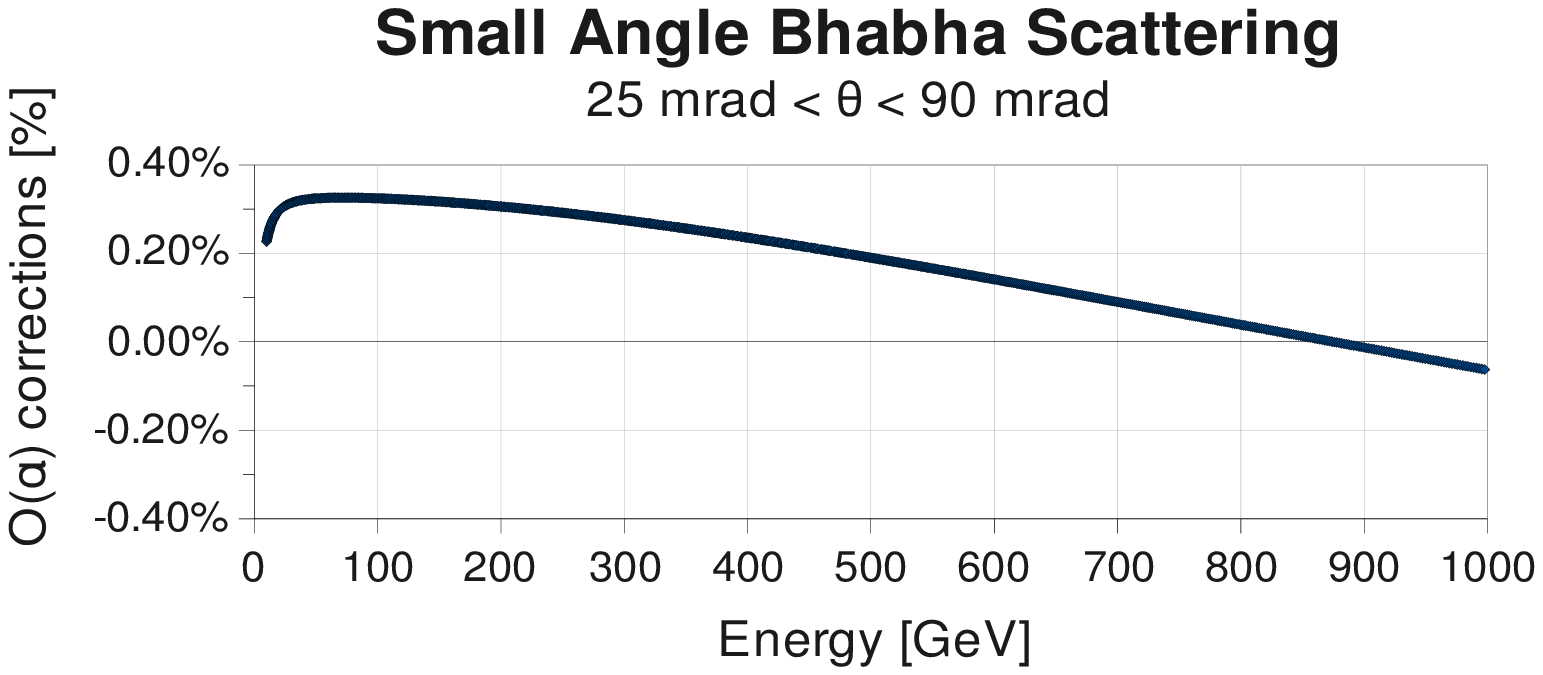}
\end{tabular}
\centering
\caption{Integrated cross section between $25~\mathrm{mrad} < \theta < 90~\mathrm{mrad}$ for Bhabha scattering (above). The percentage of the one-loop corrected cross section with respect to the tree-level is also shown (below), with a maximum of $0.33\%$ at \mbox{$E=m_Z$}. The maximum of soft photon energy was taken to be $E^\mathrm{max}_{\gamma_\mathrm{soft}}=0.2 \sqrt{s}$.}
\label{sabh}
\end{figure}

Fig.~\ref{sabh} depictes the second configuration for small angle scattering in the forward region and the relative importance of the first order ($\mathcal{O}(\alpha)$) perturbative corrections. Here the resonance disappears due to the large contributions coming from the divergent Feynman diagram exchanging a photon at the collinear case. In this case the relative corrections are small, about the permill order: $\mathcal{O}(10^{-3})$. As mentioned before, this study should be supplemented with the hard-photon bremmstrahlung and second order perturbative corrections in order to specify the amount of uncertainty coming from the theoretical prediction.


\section{Conclusions and outlook}
\label{conclusions}
Porting the application was successful with the help of the \gridway{} metascheduler. A scheme of master/worker was developed where the master remain at user's side, giving instructions through \gridway{} about how the submission should be partitioned and managed. Because the code created by \aitalc{} is a dynamically linked executable, a few modifications to the original code were required in order to safely run as a worker under different configurations at every Grid node. Different workload balances were studied for the sake of performance, finding out a reasonable default behaviour without any configuration tweaking. Still the time of waiting for a working node limites improvements when making the jobs smaller. Therefore submitting jobs which require less than a few minutes to complete do not increase performance because Grid latencies start playing a significant role.

Future work includes fine-graining of data results through feedback into the master and standardization of porting of similar codes developed without parallel execution in mind. An interesting possibility would be the implementation of more advanced job managers which dynamically obtain feedback from finished jobs and try strategies according to well defined policies. This could minimize waiting time, avoiding strictly failing nodes or avoiding resubmission of jobs being processed slowly.

\section*{Acknowledgments}
The authors would like to thank the phenogrid VO for letting us access their resources for this study. We also welcome the openness of this  UK/Ireland regional declared VO to other European partners and users interested in the phenomenology of particle physics.

This work makes use of results produced with the \href{http://www.eu-egee.org}{EGEE} grid infrastructure, co-funded by the European Commission (INFSO-RI-222667).

This research was supported by Consejería de Educación de la Comunidad
de Madrid, Fondo Europeo de Desarrollo Regional (FEDER) and
Fondo Social Europeo (FSE), through BIOGRIDNET Research Program
S-0505/TIC/000101, by Ministerio de Educación y Ciencia, and through
the research grant TIN2006-02806, and by the European Union through
the research grant EGEE-III grant agreement 22667.

\end{document}